\documentclass[superscriptaddress,floatfix,%
 aip,
 amsmath, %amssymb,
reprint,%
]{revtex4-1}

\usepackage{graphicx}% Include figure files
\usepackage{dcolumn}% Align table columns on decimal point
\usepackage{bm}% bold math
\usepackage[utf8]{inputenc}
\usepackage[T1]{fontenc}
\usepackage{mathptmx}
\usepackage{etoolbox}

\usepackage[group-separator = {,}]{siunitx}
\usepackage[normalem]{ulem}  % to use \sout{}

\DeclareSIUnit[per-mode = symbol]\density{\unit{\gram\per\cm^3}}
\DeclareSIUnit[per-mode = symbol]\effdose{\unit{\pico\sievert\;\cm^2}}
\DeclareSIUnit[per-mode = symbol]\wt{\unit{wt\%}}
\DeclareSIUnit[per-mode = symbol]\barn{b}
\usepackage{array,multirow}
\usepackage{hyperref} % load close to last to minimize package clashes
\usepackage{subfiles} % Best loaded last in the preamble

\newcolumntype{L}[1]{>{\raggedright\let\newline\\\arraybackslash\hspace{0pt}}m{#1}}
\newcolumntype{R}[1]{>{\raggedleft\let\newline\\\arraybackslash\hspace{0pt}}m{#1}}
\newcolumntype{C}[1]{>{\centering\let\newline\\\arraybackslash\hspace{0pt}}m{#1}}

\begin{document}

%\nolinenumbers
\preprint{AIP/123-QED}

\title{Designing a boron nitride polyethylene composite for shielding neutrons} %Title of paper



\author{A. D. Vira}
\email[avira@gatech.edu]{avira@gatech.edu}
%\thanks{}
\affiliation{School of Physics, Georgia Institute of Technology, Atlanta, GA 30332, USA\looseness=-1}

\author{E. M. Mone}
\affiliation{School of Physics, Georgia Institute of Technology, Atlanta, GA 30332, USA\looseness=-1}

\author{E. A. Ryan}
\affiliation{School of Materials Science and Engineering, Georgia Institute of Technology, Atlanta, GA 30332, USA\looseness=-1}

\author{P. T. Connolly}
\affiliation{George W. Woodruff School of Mechanical Engineering, Georgia Institute of Technology, Atlanta, GA 30332, USA\looseness=-1}

\author{K. Smith}
\affiliation{Space Science and Applications (ISR-1), Los Alamos National Laboratory, Los Alamos, NM 87545, USA\looseness=-1}

\author{C. D. Roecker}
\affiliation{Space Science and Applications (ISR-1), Los Alamos National Laboratory, Los Alamos, NM 87545, USA\looseness=-1}

\author{K. E. Mesick}
\affiliation{Space Science and Applications (ISR-1), Los Alamos National Laboratory, Los Alamos, NM 87545, USA\looseness=-1}

\author{\\T. M. Orlando}
\affiliation{School of Physics, Georgia Institute of Technology, Atlanta, GA 30332, USA\looseness=-1}
\affiliation{School of Chemistry and Biochemistry, Georgia Institute of Technology, Atlanta, GA 30332, USA\looseness=-1}

\author{Z. Jiang}
\affiliation{School of Physics, Georgia Institute of Technology, Atlanta, GA 30332, USA\looseness=-1}

\author{P. N. First}
\email[first@gatech.edu ]{first@gatech.edu}
\affiliation{School of Physics, Georgia Institute of Technology, Atlanta, GA 30332, USA\looseness=-1}

\date{\today}

\begin{abstract}
Neutrons are encountered in many different fields, including condensed matter physics, space exploration, nuclear power, and healthcare. Neutrons interacting with a biological target produce secondary charged particles that are damaging to human health. The most effective way to shield neutrons is to slow them to thermal energies and then capture the thermalized neutrons. These factors lead us to consider potential materials solutions for neutron shields that maximize the protection of humans while minimizing the shield mass, and which adapt well to modern additive manufacturing techniques. Using hexagonal boron nitride (hBN) as a capture medium and high-density polyethylene (HDPE) as a thermalization medium, we aim to design the optimal internal structure of h$^{10}$BN/HDPE composites by minimizing the effective dose, which is a measure of the estimated radiation damage exposure for a human. Through Monte Carlo simulations in Geant4, we find that the optimal structure reduces the effective dose up to a factor of 72x over aluminum (Al) and 4x over HDPE; this is a significant improvement in shielding effectiveness that could dramatically reduce the radiation exposure of occupational workers.
\end{abstract}

\maketitle

%% ------------------------------------------------------%%
%\linenumbers
\section{Introduction}\label{sec:introduction}

Neutron radiation---encountered in healthcare, nuclear reactors, and the space environment---is extremely damaging to human health \cite{lehmann2017recent,goldsmith1989mortality,townsend2005implications}. Long-term exposure results in cancer, cognitive decline, and degeneration of the circulatory system \cite{united1996sources,valentin2007ICRP}. Neutrons interacting with a biological target produce secondary charged particles, which have a high linear energy transfer (LET, keV/$\mu$m) causing subsequent tissue damage \cite{baiocco2016origin}. Shielding against neutron radiation is difficult since neutrons can only be moderated or absorbed. The scattering cross-sections of neutron captures are typically orders of magnitude smaller at high energies ($> \SI{1}{\MeV}$) so the neutrons need to be moderated. In addition, thermal neutrons produce damaging secondary radiation through the neutron capture reaction that typically emits prompt gammas, which are also damaging to health, so ultimately the thermal neutrons should be absorbed to eliminate the danger.

In this work, we consider the physical processes of moderation and absorption to address a basic question: To protect human health, what is the optimal distribution of thermalization and capture elements within a shielding material? Prior experimental studies have considered different internal distributions of materials (e.g. continuous blends or layered composites) for a limited number of configurations, finding an improvement in some configurations \cite{shang2020multilayer,shin2014polyethylene,yang2020polymer,harrison2008polyethylene,maity2016realization,zhang2017enhancing}. By addressing this question from a computational perspective using extensive Monte Carlo simulations, we explore a larger configuration space, including structures that are appropriate for modern manufacturing techniques, such as additive manufacturing (AM). The result of this work has clear applications. For example, conventional high-mass radiation shielding is employed for many terrestrial applications but would be a poor choice for space environments such as the Moon, due to associated launch costs. Lightweight polymer materials and composites are increasingly employed for neutron shielding applications and are compatible with a variety of AM techniques \cite{sacco2019additive}. AM capabilities have grown rapidly in recent years and are planned for autonomous lunar construction \cite{clinton2021overview}. The shielding composites, explored in this study, would be of particular interest for lunar applications given the lightweight materials and compatibility with AM. 

Because the capture cross-section scales inversely with velocity, reducing the neutron energy improves the capture efficiency. To accomplish these two physical processes, moderation and absorption, we select appropriate materials. Effective moderation requires a high rate of momentum transfer to the material, which is best accomplished by compounds with high hydrogen content due to the near equality of neutron and proton masses. A good moderator is capable of slowing down fast neutrons without undergoing significant radiation damage or capturing the neutrons. Water ($\rho \approx \SI{1}{\density}$), graphite ($\rho \approx \SI{2.26}{\density}$), and polyethylene ($\rho \approx$ 0.90 – $\SI{0.97}{\density}$) are commonly-chosen moderator materials. High-density polyethylene (HDPE), (C$_2$H$_4$)$_n$ for $n$ between 3,500 and 9,000, is ideal for a variety of applications due to its low density and large elastic neutron scattering cross-section. Additionally, unlike water and graphite, HDPE has suitable physical properties to be used as a stand-alone flexible and light-weight radiation shielding material across a broad range of operating temperatures, which is potentially useful for shielding neutrons on the lunar surface \cite{mesick2018benchmarking}. 

Among low mass isotopes, $^{10}$B, comprising 19.9\% of natural boron, is the clear champion for neutron capture with a cross-section that reaches $\SI{3800}{\barn}$ for thermal neutrons \cite{plompen2020joint}. The following capture reactions are relevant for $^{10}$B:
\begin{subequations}
\label{eq:whole}
\begin{equation}
{}^{10}_5 \text{B} + {}^1_0\text{n}_{th} \rightarrow {}^{7}_3 \text{Li}^* \text{ ($\SI{0.84}{MeV}$)} + {}^{4}_2 \text{$\alpha$}^* \text{ ($\SI{1.47}{MeV}$)},\label{subeq:1}
\end{equation}
\begin{eqnarray}
{}^{10}_5 \text{B} + {}^1_0\text{n}_{th} \rightarrow {}^{7}_3 \text{Li} \text{ ($\SI{1.01}{MeV}$)} + {}^{4}_2 \text{$\alpha$} \text{ ($\SI{1.78}{MeV}$)}.\label{subeq:2}
\end{eqnarray}
\end{subequations}
Here, $^{10}$B captures the thermal neutrons (${}^1_0\text{n}_{th}$), converting them into alpha particles ($\alpha$) and lithium ions, which have a stopping range of only a few micrometers at the relevant energies---much shorter than the neutron attenuation length. The neutron capture reaction of Equation~\ref{subeq:1}, with emission of a $\SI{0.478}{MeV}$ gamma ray, has a $\sim$94\% probability and the reaction of Equation~\ref{subeq:2} has a $\sim$6\% probability.

Prior experimental studies of neutron transmission through boron-containing HDPE composites reveal an improvement in the shielding effectiveness in comparison to pure HDPE \cite{shang2020multilayer,zhang2017enhancing,shang2020multilayer,yang2020polymer,harrison2008polyethylene,maity2016realization}. A high boron-containing material is more efficient at capturing neutrons. B$_4$C has a higher boron content than hexagonal boron nitride (hBN); however, Shin et al., 2014 showed that hBN can be functionalized to improve the solubility and homogeneity of the composite, which enhanced its shielding effectiveness beyond that of B$_4$C composites \cite{shin2014polyethylene}. Hexagonal boron nitride, a quasi-2D material with a large bandgap of $\sim\SI{5.9}{eV}$, has naturally occurring boron (19.9\% of ${}^{10}$B and 80.1\% of ${}^{11}$B, denoted as h${}^{\text{Na}}$BN) and can undergo the capture events shown in Equations~\ref{subeq:1}-\ref{subeq:2}. However, hBN can be boron-10 enriched (denoted as h${}^{10}$BN), further increasing the probability of undergoing the capture reaction \cite{liu2018single}. This work provides a comprehensive characterization that directly builds upon previously published experimental works \cite{shang2020multilayer,zhang2017enhancing,shang2020multilayer,yang2020polymer,harrison2008polyethylene,maity2016realization} by using an open-source Monte Carlo (MC) code, Geant4 (GEometry ANd Tracking), to optimize the shielding composition and structures for h${}^{10}$BN/HDPE composites \cite{agostinelli2003geant4,allison2006geant4,allison2016recent}. Geant4 is a widely used and validated Monte Carlo code that offers freedom in creating the detector geometry and a range of physics models appropriate to different situations. In the context of our own research program, the choice of Geant4 was made because it also has the potential to incorporate detailed solid-state effects.

For human safety, we aim to maximize shielding effectiveness by optimizing the boron distribution within polyethylene composites using the effective dose as a figure of merit. The effective dose is a quantity that provides radiation workers with an assessment of the potential hazard to human health by accounting for the radiation effects on human organs and tissue. Defining the effective dose ratio as the effective dose with a shield divided by the effective dose without a shield, we run Geant4 simulations for three types of composites: (1) homogenous blend of HDPE and hBN (“blended”), (2) pure HDPE interlayered with pure hBN (“ideal layered”, inspired by \cite{sazali2019review}), and (3) pure HDPE interlayered with an hBN/HDPE blended composite (“manufacturable”). For all configurations, the Geant4 simulations assume a homogeneous blend of the specified elements and do not consider the impact of component particle sizes or shapes, which could affect the shielding efficiency \cite{cheewasukhanont2020effect,kim2021analysisshielding}. The influences of particle size, geometry, and distribution are topics for future research; within our present approximation, we expect that the simulations should remain valid for composites with inhomogeneities on a size scale much smaller than the neutron-capture mean-free path, which in h${}^{10}$BN is $\sim\SI{10}{\micro\meter}$ for thermal neutrons. The manufacturable configuration could be fabricated via multi-material fused filament fabrication, an AM technique where thermoplastic filaments are deposited to build up 3D geometries \cite{wang2020dispersion,sacco2019additive}. The simulated incident neutrons have a log-uniform energy distribution from $\SI{15}{meV}$ to $\SI{20}{MeV}$. During post-processing, we separate the neutrons by different energy regimes, discussed in Section \ref{subsec:subsection_postprocessing}, to isolate the physics occurring within each regime \cite{carron2006introduction}. This also widens the applications of hBN/HDPE shielding composites by providing results that can be applied generally to shielding neutrons for an arbitrary distribution of incident energies. We find that alternating layers of capture and moderation materials provide improved radiation protection compared to blended composites with the same composition. From the manufacturable configurations, we find that we can improve the shielding effectiveness up to a factor of 72 relative to Al and up to a factor of 4 with respect to HDPE. This is a significant improvement that could dramatically reduce the occupational radiation dose for workers in high-risk environments, for example, astronauts or healthcare professionals. 

Section \ref{sec:FoM} discusses the figure of merit that is used to evaluate the shielding effectiveness of all the simulated shielding composites. The structure of the Geant4 simulations is discussed in Section \ref{sec:structure}. We conclude the paper by discussing the results for the three different geometries for the hBN/HDPE composites (blended, ideal layered, and manufacturable) and the corresponding optimal configurations in Sections \ref{sec:results} and \ref{sec:discussion}. 

%% ------------------------------------------------------%%
\section{Figure of Merit: Effective Dose}\label{sec:FoM}

There are three dose quantities, in radiation physics, that are used to assess radiation exposure: absorbed dose, equivalent dose, and effective dose. The absorbed dose is defined as the mean energy imparted by ionizing radiation in matter per unit of mass (the SI unit of absorbed dose is the Gray, with $\SI{1}{\gray} = \SI{1}{\joule\per\kilogram}$). The equivalent dose ($H_T$) weights the absorbed dose by the radiation type to determine the effect of the ionizing radiation on a specific organ or tissue. The effective dose ($E$), provided in units of Sieverts (Sv), takes the equivalent dose and sums over the organ/tissue. Mathematically, the effective dose reads
\begin{equation} \label{eq:2}
E = \sum_T w_T H_T = \sum_{T,R} w_T w_R D_{T,R}
\end{equation}
where $w_R$ is the radiation weighting factor, $w_T$ is the tissue weighting factor, and $D_{T,R}$ is the mean absorbed dose from a radiation type, $R$, in specific organs or tissues, $T$. 

The International Commission on Radiological Protection (ICRP) provides a set of protection quantities based on radiation transport simulations for a monoenergetic particle emitted in various directions (rotational, left lateral, right lateral, isotropic, etc.) incident on a human phantom. ICRP provides the necessary coefficients to convert incident energy to effective dose for sex-averaged models of human phantoms. These coefficients are used to estimate the effective dose an average human receives for a given radiation type and energy. Here, we use the conversion coefficients for an isotropic input distribution as it more accurately represents the radiation in space. ICRP 103 provides the radiation weighting factors \cite{valentin2007ICRP}; ICRP 116 and 123 provide the conversion coefficients for a given fluence ($\Phi$) \cite{petoussi2010ICRP, petoussi2014ICRP}.

\begin{figure}
\includegraphics[width=1\columnwidth]{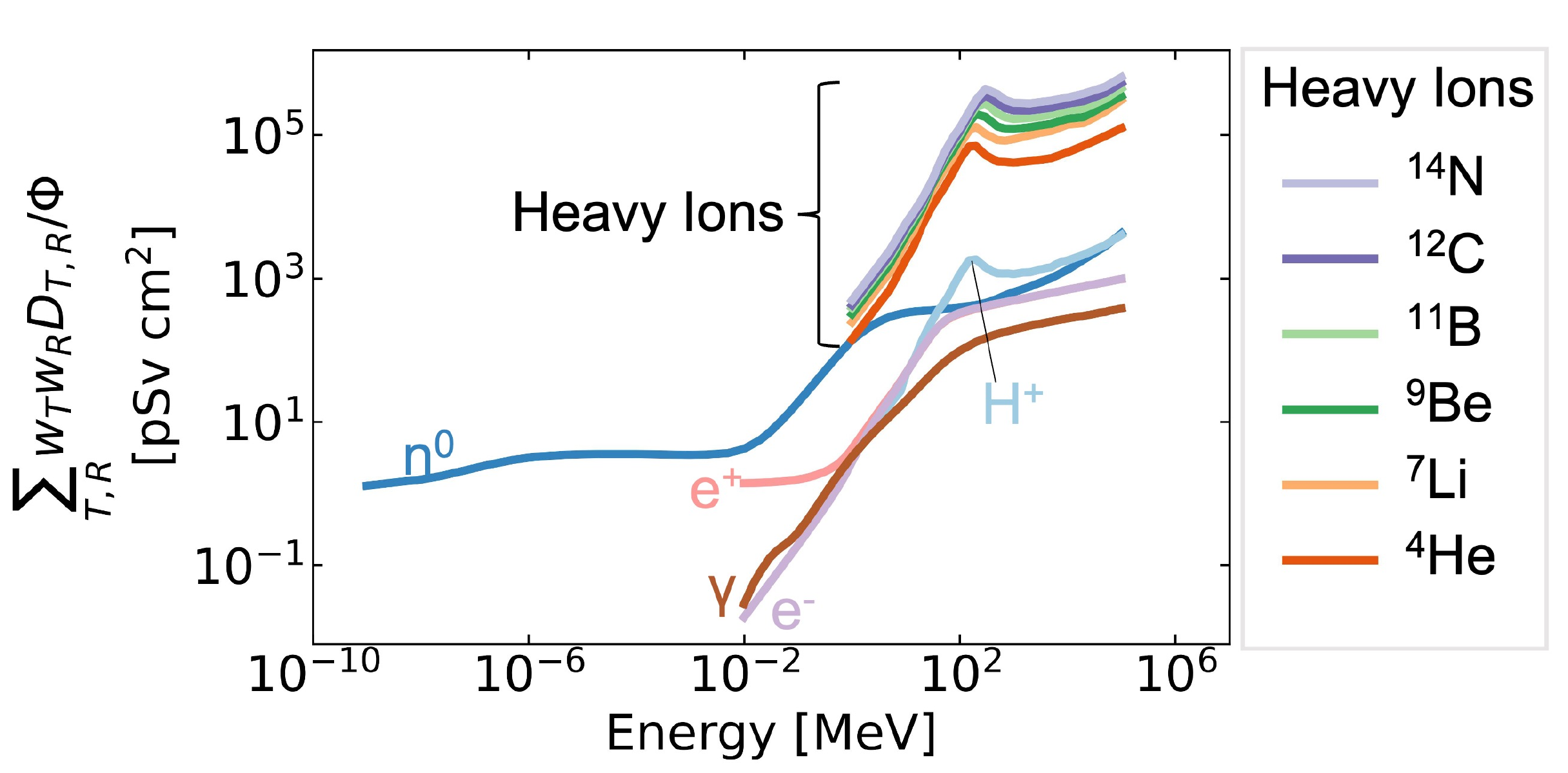}% Here is how to import EPS art
\caption{\label{fig:effective_dose} Effective dose per fluence (\si{\effdose}) for a sex-averaged human versus the incident energy, calculated for various radiation types emitted isotropically. ICRP 116 and 123 are used to obtain the conversion factors from energy to effective dose \cite{petoussi2010ICRP,petoussi2014ICRP}}
\end{figure}  

Figure~\ref{fig:effective_dose} shows the effective dose per fluence ($E/\Phi$) as a function of the incident energy for different radiation types. The heavy ions have the largest effective dose and can, therefore, be the most damaging for humans compared to other ionizing particles. Gamma rays, positrons, electrons, and protons have a similar effective dose and are also important to shield against, especially for energies above $\sim \SI{1}{\MeV}$. Neutrons are damaging across all incident energies, showing the importance of incorporating $^{10}$B into the shielding components. 

%% ADV: CHANGE WEIGHT PERCENTAGE TO MASS FRACTION IN SOME PLACES ???

%% ------------------------------------------------------%%
\section{Structure of Geant4 Simulations}\label{sec:structure}

\subsection{Design Space for hBN/HDPE Composites}

We restrict our design space to address the use of h${}^{10}$BN and HDPE to perform the two-step shielding process, neutron  capture and thermalization, respectively. Our goal is to explore the design space (blended, ideal layered, and manufacturable) of h${}^{10}$BN/HDPE composites using the effective dose as a figure of merit to obtain the optimal boron distribution in HDPE composites. 

\paragraph*{Blended Geometry.}
For the blended geometry, two free design parameters are considered: total composite thickness ($t$) and global weight percentage of h${}^{10}$BN ($\omega_{global}$). The thickness is varied between $\SI{1}{\mm}$ and $\SI{150}{\mm}$ (specifically, 1,3,5,7,10,20,30,60,100,$\SI{150}{\mm}$) to simulate a shielding material for various applications, such as integration into personal protective equipment or lining for habitat shielding. The global weight percentage, $\omega_{global}$, is varied between 0 and 100\%, in increments of 10\%. The range of weight percentage is simulated to cover the entire design space of all possible blended configurations even though blended composites with $>\SI{30}{\wt}$ of h${}^{10}$BN are difficult to manufacture via traditional blending methods due to poor particle dispersion and dramatically increased viscosity of the melt at higher loadings, particularly when hBN nanoparticles are used in the manufacture of blended composites \cite{wei2022study,lebedev2022comparative,wang2020dispersion}. Including higher weight percentages of h${}^{10}$BN, beyond the range of manufacturable composites, also allows this dataset to be used to evaluate the cost-benefit relationship for the implementation of advanced manufacturing methods which could be used to produce composites with higher weight percentages of h${}^{10}$BN. For each thickness, a set of 13 simulations are run to cover the design space of $\omega_{global}$, resulting in 130 unique simulations for the blended geometry. 

\paragraph*{Ideal Layered Geometry.}
For the ideal layered geometry, we consider three parameters: total composite thickness ($t$), number of HDPE and h${}^{10}$BN layers within a composite thickness (where $n$ is specifically the number of h${}^{10}$BN layers), and global weight percentage of h${}^{10}$BN ($\omega_{global}$). This configuration consists of interlayering pure HDPE and pure h${}^{10}$BN layers to explore how spatial separation affects the effective dose compared to a blended structure. We explore this parameter space of interlayering pure h${}^{10}$BN to evaluate if there is a benefit to the spatial separation, even though it is difficult to make a pure hBN layer of large size due to the restricted growth conditions required to obtain a pure hBN crystal \cite{watanabe2004direct}. From the capture reaction in Equations~\ref{subeq:1}-\ref{subeq:2}, it is best to terminate the layered composite with an HDPE layer, which is the same thickness as the other pure HDPE layers, so that the heavy charged particles—which would significantly increase the effective dose (see Figure~\ref{fig:effective_dose})—can be stopped before exiting the composite shielding. For this reason, there is always one less h${}^{10}$BN layer than HDPE layers. The number of h${}^{10}$BN layers is varied between 1 (2 layers of HDPE) and 99 layers (100 layers of HDPE) where the number of layers is equivalent to the number of periods. The thickness is varied over the same range of thicknesses used for the blended geometry. The global h${}^{10}$BN weight percentage, $\omega_{global}$, changes the relative thicknesses of the h${}^{10}$BN layers within the internal structure and varied between 5-\SI{95}{\wt} to cover the entire parameter space (where $\omega_{global} = 0$ is a pure HDPE block and $\omega_{global}=100$ is a pure h${}^{10}$BN block). Based on this design space, there are 1210 unique simulations (10 thicknesses, 11 different $\omega_{global}$, and 11 values of $n$) for the ideal layered geometry. 

\paragraph*{Manufacturable Geometry.}
Due to manufacturability constraints of creating thick continuous hBN layers bonded to HDPE layers in the ideal layered structure and high loadings in some blended structures, it is difficult, or impossible in some cases, to manufacture all the geometries that are explored in the blended and ideal layered geometries \cite{watanabe2004direct}. To address this issue, additional simulations were run in which pure HDPE was interlayered with h${}^{10}$BN/HPDE blended composite layers with $\leq \SI{20}{\wt}$ of h${}^{10}$BN. A weight percentage of $\leq$ 20\% is chosen as prior experimental studies have shown little change in mechanical or rheological properties of hBN/HDPE composites at these loadings \cite{wang2020dispersion}. Additionally, given the minimal influence of $\leq \SI{20}{\wt}$ on the rheological properties of hBN/HDPE composites, it is reasonable to infer that these manufacturable geometries can be adapted for fused filament fabrication, an additive manufacturing (AM) process \cite{sacco2019additive}. Using common dual extrusion technologies, the blended feedstocks, comprising of hBN nanoplatelets blended into HDPE, are interlayered with pure HDPE feedstocks to build up the 3D geometry where the layer thickness is easily controlled by simple programming. Alternatively, these materials could be manufactured into layered films via common film forming and lamination processes to create large surface area composite sheets suitable for various applications, such as habitat construction for lunar missions. 

For the manufacturable geometry, there are four design parameters: total thickness of composite material ($t$), weight percentage of the blended h${}^{10}$BN/HDPE layer ($\omega_{global}$), local weight percentage of h${}^{10}$BN within the blended layer ($\omega_{local}$), and number of mixture layers ($n$). Refer to Figure~S1, in the supplemental material, for a plot of the interdependence of these design parameters on the total areal density. For a subset of simulations, we explore the decrease in shielding effectiveness by including h${}^{\text{Na}}$BN instead of h${}^{10}$BN, which is cheaper and more available \cite{liu2018single}. The thickness of each mixture h${}^{10}$BN/HDPE layer ($t_{mix}$) is
\begin{eqnarray}
t_{mix} =&& \frac{t}{n}\left[\omega_{global} \frac{\rho_{total}}{\rho_{mix}}\right]\label{eq:3} \\
=&& \frac{t}{n}\left[\frac{\omega_{global}(1-\omega_{global})/\rho_{HDPE}+\omega_{global}^2/\rho_{mix}}{(1-\omega_{local})/\rho_{HDPE}+\omega_{local}/\rho_{hBN}}\right]\nonumber
\end{eqnarray}
where $\rho_{total}$ is the total density of the composite material, and $\rho_{mix}$ is the density of the mixture layer. The thickness of each HDPE layer ($t_{HDPE}$) is
\begin{equation} \label{eq:4}
t_{HDPE} = \frac{1}{n+1}\left[t - (n-1)t_{mix}\right].
\end{equation}

Equations~\ref{eq:3} and \ref{eq:4} can be reduced to the ideal layered geometry by assuming $\omega_{local}$ is 100\% and $\omega_{global}$ controls the relative thickness between the pure HDPE and pure h${}^{10}$BN layers. For the manufacturable configurations, $t$ is varied between $\SI{1}{\mm}$ and $\SI{150}{\mm}$ using the same thicknesses in the blended configuration, $\omega_{global}$ is varied between \SI{10}{\wt} and \SI{95}{\wt}, $n$ is varied between 1 and 9 in increments of one, and $\omega_{local}$ is varied between \SI{1}{\wt} and \SI{20}{\wt}. Based on the large design space, we run 5,670 simulations for the manufacturable configuration alone, which produces a few terabytes of data (see Appendix \ref{app:datareduction} for a description of the data reduction).

\paragraph*{Reference Geometry.}
Simulations of pure aluminum, lead, and HDPE block are carried out to provide a direct comparison with commonly used radiation shielding materials. These simulations are run for similar areal densities of the h${}^{10}$BN/HDPE composites to allow for a direct comparison. 

\subsection{Implementation of Simulations}

Five million neutrons, isotropic in angle, are uniformly incident on one side of the composite material with energies randomly sampled from $\SI{15}{meV}$ to $\SI{20}{MeV}$ with a log-uniform distribution. Within each Geant4 simulation, a “sensitive detector” is placed on the entry and exit surface of each composite structure to record the particles entering and exiting the material without saving all the interactions occurring within it (see Appendix \ref{app:geant4details} for details of Geant4). The composite material is set to have periodic boundary conditions (using the g4pbc module) in the lateral dimensions to simulate an infinite slab. The lateral dimension for all the simulations is $\SI{100}{\cm}$, which is $\sim$10x larger than the thickest sample. We choose such a large lateral dimension to ensure that the periodic boundary conditions do not have a significant effect on the simulation results. %Note that an equivalent method is the use of isotropically emitted neutrons from a point source and a large lateral dimension to ensure that the particles are not escaping through the lateral edges. %We also note this could potentially be done more efficiently by running one simulation at the largest thickness, namely $\SI{150}{\mm}$, and extract the exiting distribution for smaller thicknesses; for simplicity, we run each thickness as a separate run to minimize the personnel time required for post-processing. 

\subsection{Post-processing of Geant4 Results}\label{subsec:subsection_postprocessing}

The incident neutrons are separated into distinct energy bins to isolate the physics driving the neutron interactions and widen potential applications of this work by allowing optimization of hBN/HDPE composite for specific energy distribution (e.g. albedo lunar neutrons or neutron reactors). The neutron energy bins, shown in Table~\ref{tab:table1}, are fairly standard bins \cite{carron2006introduction}. Within each energy range, the physics behind the dominant neutron interactions leads to their names.

In thermal equilibrium, neutrons approach a Maxwell-Boltzman distribution with the peak near $\SI{0.025}{\eV}$. Neutrons are referred to as cold neutrons below the peak, thermal neutrons at the peak, and epithermal neutrons above the peak. At an energy between $\SI{0.4}{\eV}$ and $\SI{0.6}{\eV}$, neutrons have a large capture cross-section with $^{113}$Cd and, therefore, referred to as cadmium neutrons. Above the cadmium energy range is referred to as epicadmium, which ends at $\SI{1}{\eV}$ since the capture cross-section drops by an order of magnitude. After the epicadmium range, neutrons from $\SI{1}{\eV}$ to $\SI{10}{\eV}$ are referred to as slow neutrons since nuclear resonance typically begins at $\SI{10}{\eV}$. Above $\SI{10}{\eV}$, there are typically resonances in the capture cross-sections. The cutoff energy for resonance neutrons is not well-defined but typically stops at $\SI{300}{\eV}$. Between $\SI{300}{\eV}$ and $\SI{1}{\MeV}$, there are nuclear reactions that produce other particles, such as protons and alpha particles; this range is called the intermediate energies. Fast neutrons cut off at an energy of $\SI{20}{\MeV}$ since the corresponding velocity is 20\% of the speed of light. To reduce the number of bins and increase the signal-to-noise ratio, we combine the following energies: thermal with cold, cadmium with epithermal, and slow with epicadmium. 

\begin{table}
\caption{\label{tab:table1}Neutron energy bins (determined from \cite{carron2006introduction}) that are used to post-process the Geant4 simulations. }
\begin{ruledtabular}
\begin{tabular}{lR{0.05\columnwidth}R{0.18\columnwidth}C{0.05\columnwidth}L{0.18\columnwidth}}
 Neutron Energy Name & \multicolumn{4}{c}{Energy Range}  \\
\hline 
 Cold (up to thermal peak) & & $\SI{15}{\meV} <$& E & $\leq$ $\SI{0.025}{\eV}$ \\
 Epithermal (includes cadmium) & & $\SI{0.025}{\eV}$ $<$ & E & $\leq$ $\SI{0.5}{\eV}$ \\
 Epicadmium (includes slow) & & $\SI{0.5}{\eV}$ $<$ & E &  $\leq$ $\SI{10}{\eV}$  \\
 Resonance & & $\SI{10}{\eV}$ $<$ & E & $\leq$ $\SI{300}{\eV}$  \\
 Intermediate & & $\SI{300}{\eV}$ $<$ & E & $\leq$ $\SI{1}{\MeV}$\\  
 Fast & & $\SI{1}{\MeV}$ $<$ & E & $\leq$ $\SI{20}{\MeV}$ \\ 
\end{tabular}
\end{ruledtabular}
\end{table}

%% ------------------------------------------------------%%
\section{Composite Simulation Results}\label{sec:results}

\subsection{Blended Geometry}

\begin{figure}
\includegraphics[width=1\columnwidth]{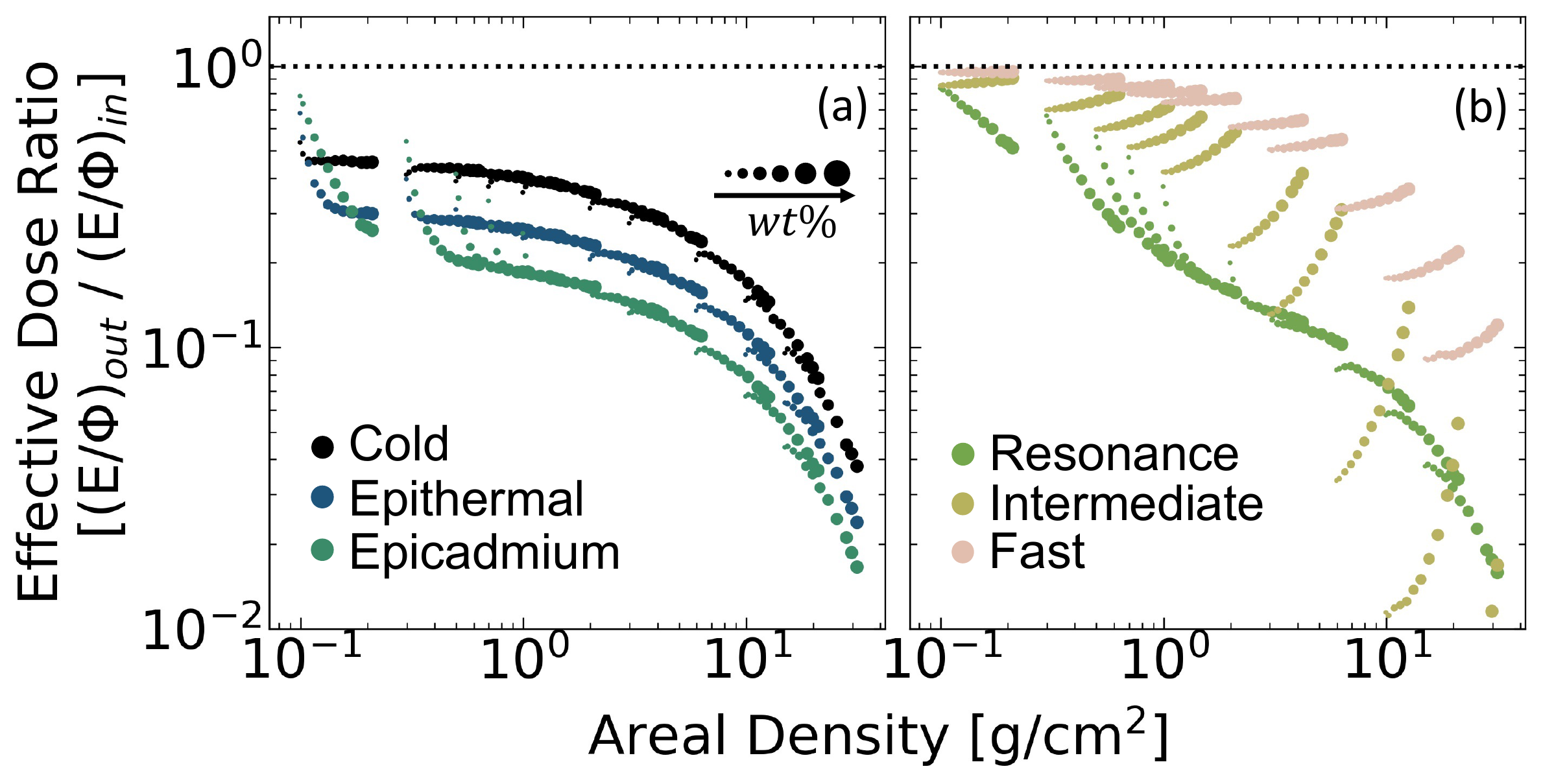}
\caption{\label{fig:blended} Effective dose ratio ($(E/\Phi)_{out}/(E/\Phi)_{in}$) versus areal density (g/cm$^2$) for blended h${}^{10}$BN/HDPE composites. In both panels, each color corresponds to a different energy range and the marker size increases with increasing $\omega_{global}$. (a) shows the results for neutron energies below $\SI{10}{\eV}$ and (b) shows the results for neutron energies above $\SI{10}{\eV}$. The optimal composite changes with areal density for energies below the intermediate range; above the intermediate range, a pure HDPE block is the most effective configuration. }
\end{figure}

We explore the shielding effectiveness of a homogeneous distribution of boron in HDPE by varying $\omega_{global}$ from 0\% to 100\%  in 10\% increments using the effective dose as a figure of merit (described in Section \ref{sec:FoM}). We optimize the effective dose ratio ($(E/\Phi)_{out}/(E/\Phi)_{in}$) by calculating the effective dose with and without the shielding material. Figure~\ref{fig:blended} shows the effective dose ratio as a function of areal density (g/cm$^2$) for all h${}^{10}$BN/HDPE blended configurations. In both panels, the marker size increases linearly as $\omega_{global}$ increases from 0\% to 100\% (the smallest marker corresponds to a pure HDPE block, largest marker corresponds to a pure h${}^{10}$BN block), and each color corresponds to a unique energy range, shown in Table~\ref{tab:table1}, based on the neutron energy entering the composite. 

In Figure~\ref{fig:blended}a, the shielding efficiencies of the h${}^{10}$BN/HDPE composites changes with the incident neutron energy and areal density. For example, in the epithermal range, the optimal structure for areal densities below $\sim \SI{2}{\gram\per\cm^2}$ is a pure h${}^{10}$BN block (corresponding to the largest marker size). This result indicates the effective dose ratio is not affected by the thermalization process and, instead, the neutron capture with ${}^{10}$B is the dominant physical process that contributes to the decrease in the effective dose. However, as the areal density increases above $\sim \SI{2}{\gram\per\cm^2}$, a pure HDPE block, the smallest marker, would maintain the lowest effective dose ratio (note that the calculations are for discrete thicknesses), showing that the optimal structure changes with areal density. Similar behavior is observed for cold, epicadmium, and resonance neutrons.

Figure~\ref{fig:blended}b shows the effective dose ratio versus areal density for resonance, intermediate, and fast neutrons. The behavior is different for intermediate and fast neutrons since the optimal structure is a pure HDPE block (smallest marker size) across all areal densities $< \SI{300}{\gram\per\cm^2}$. This indicates that the effective dose ratio is reduced the most when all of the available material is devoted to theramlization because: (1) cold and thermal neutrons contribute less to the overall effective dose than more energetic neutrons (Figure~\ref{fig:effective_dose}) and (2) at lower neutron energies the probability of undergoing a capture event increases since the capture cross-section is inversely proportional to velocity.

\subsection{Ideal Layered Geometry}

\begin{figure}
\includegraphics[width=1\columnwidth]{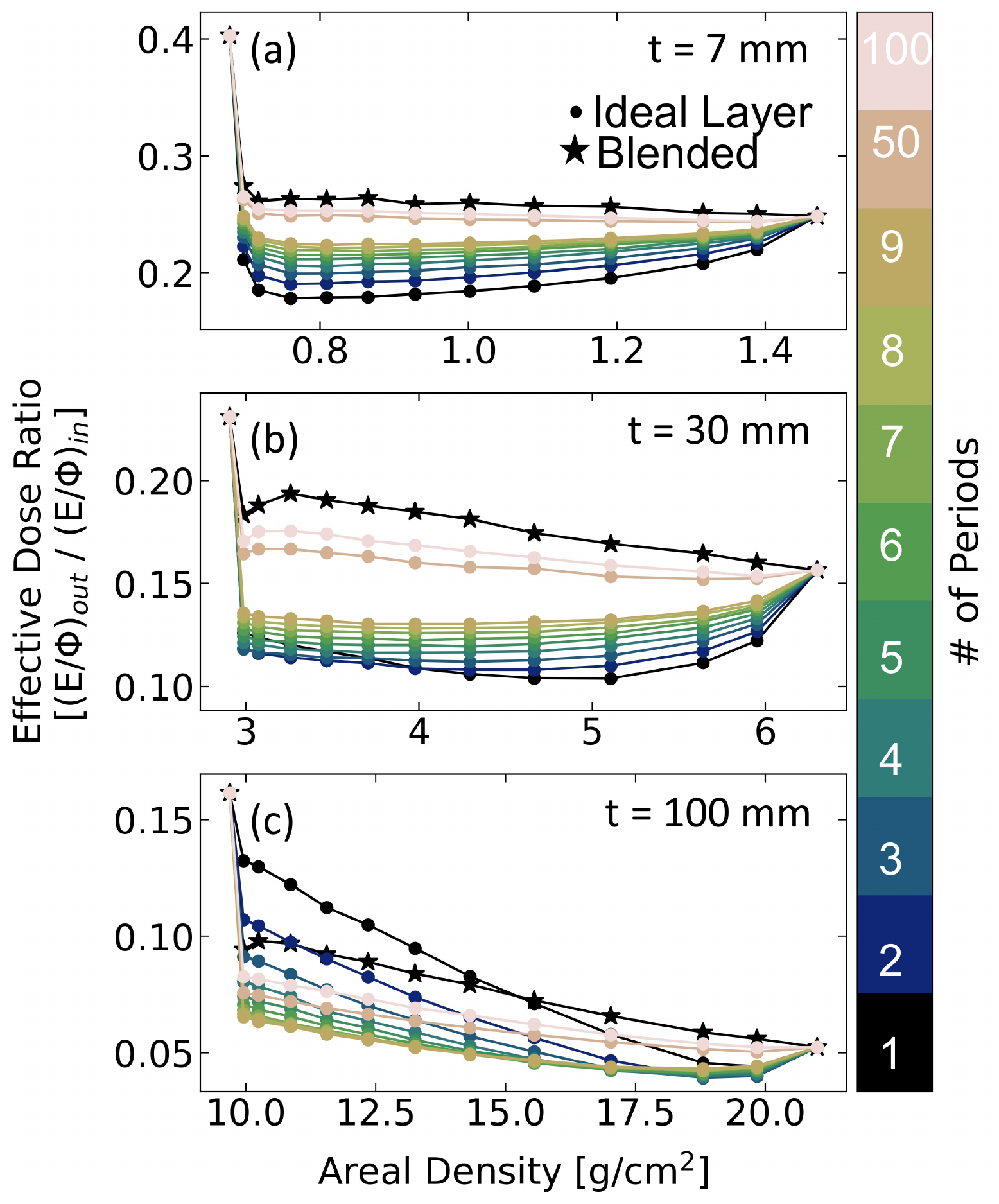}
\caption{\label{fig:layered} Epithermal Range: Effective dose ratio versus areal density for layered h${}^{10}$BN/HDPE composites with a total thickness of $\SI{7}{\mm}$ in (a), $\SI{30}{\mm}$ in (b), and $\SI{100}{\mm}$ in (c). In all panels, the color corresponds to a different number of periods, as shown in the colorbar. The blended results are shown in black stars. The ideal layered structure reduces the effective dose ratio more than the blended composite. In addition, the optimal configuration changes from a trilayer structure to a multilayer structure as the total thickness increases to $\SI{100}{\mm}$. }
\end{figure}

To further investigate the role of internal structure, we consider the next simplest modification, material layering, which separates the two essential processes of neutron thermalization and neutron capture into distinct repeating regions. We first consider a complete separation, where the layers are either HDPE or h${}^{10}$BN — a situation we refer to as the ideal layering. Figure~\ref{fig:layered} shows the effective dose ratio as a function of areal density, within the epithermal energy range, for three example composite thicknesses: $\SI{7}{\mm}$ in (a), $\SI{30}{\mm}$ in (b), and $\SI{100}{\mm}$ in (c). The different colors correspond to a unique number of h${}^{10}$BN layers, or equivalently, the spatial period of the repeating structure. As a reference, the blended results are duplicated from Figure~\ref{fig:blended} (shown in black stars). The individual thicknesses of the pure HDPE and h${}^{10}$BN layers can be calculated using Equations~\ref{eq:3} and \ref{eq:4} by setting $\omega_{local}=100$\% where $t_{mix}$ would, instead, refer to the thickness of pure h${}^{10}$BN.

From Figure~\ref{fig:layered}, we find the ideal layered geometry provides better shielding than a homogeneous blend. For $t = \SI{7}{\mm}$, the one period trilayer structure (h$^10$BN between two HDPE layers) minimizes the effective dose across all areal densities and results converge to the blended configuration as the number of layers reaches 50+. However, for $t = \SI{30}{\mm}$, the trilayer configuration minimizes the effective dose ratio only above $\sim\SI{4}{\gram\per\cm^2}$. At $t = \SI{100}{\mm}$, the optimal structure has six to eight periods versus one period for $t = \SI{7}{\mm}$, as shown in Figure~\ref{fig:layered}c and \ref{fig:layered}a. This is the first indication that the spatial scale and organization of thermalization and capture materials within a composite shield can significantly affect the macroscopic shielding properties.

\subsection{Manufacturable Geometry}\label{sec:manufacturable_results}

\begin{figure}
\includegraphics[width=1\columnwidth]{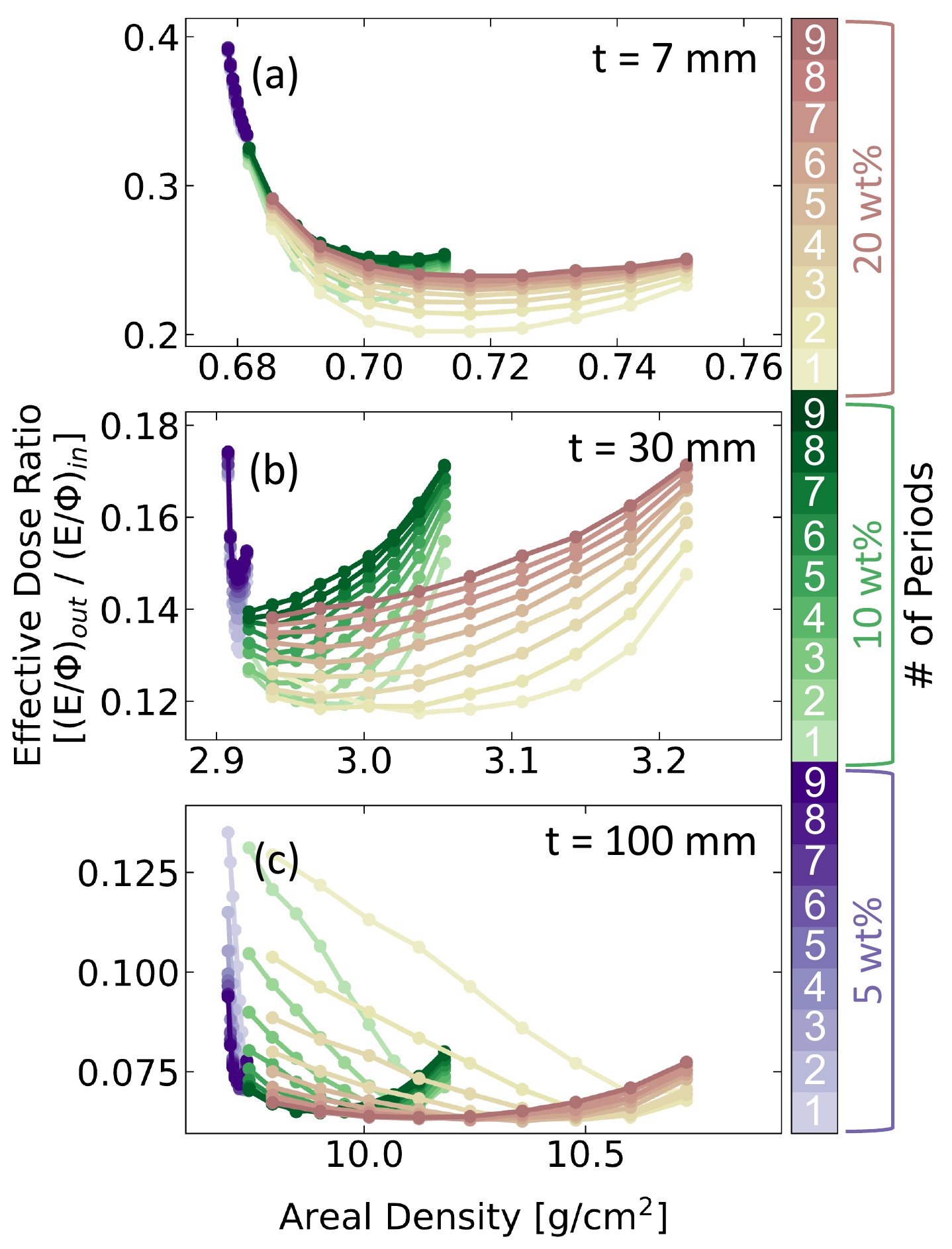}
\caption{\label{fig:manufacturable} Epithermal Range: Effective dose ratio versus areal density for layered manufacturable composites with a total thickness of $\SI{7}{\mm}$ in (a), $\SI{30}{\mm}$ in (b), and $\SI{100}{\mm}$ in (c). The manufacturable data in each panel are split into three categories, $\omega_{local}$ = 5\% (shades of brown), $\omega_{local}$ = 10\% (shades of green), and $\omega_{local}$ = 20\% (shades of purple). The different shades correspond to the different numbers of mixture layers or the spatial period of the repeating structure. For each thickness, the largest loading of h${}^{10}$BN of $\omega_{local}$ = 20\% reduces the effective dose ratio the most.} 
\end{figure} 

Given that the ideal layered geometry reduces the effective dose ratio, we explore a design space that is feasible to manufacture, as described in Section \ref{sec:structure}. Figure~\ref{fig:manufacturable} shows the effective dose ratio versus areal density for the manufacturable configuration within the epithermal energy range. Figures \ref{fig:manufacturable}a-\ref{fig:manufacturable}c show the results for $t = \SI{7}{\mm}$, $t = \SI{30}{\mm}$, and $t = \SI{100}{\mm}$, respectively (similar to Figure \ref{fig:layered}). The color bar in Figure \ref{fig:manufacturable} is split into three sections that show the $\omega_{local}$ at 5\% (shades of brown), 10\% (shades of green), and 20\% (shades of purple). The different shades of each color correspond to the number of h${}^{10}$BN/HDPE mixture layers between pure HDPE layers. For the manufacturable configurations, the optimal configuration occurs at $\omega_{local}$ = 20\% (shown in the shades of purple) for each thickness shown in Figure~\ref{fig:manufacturable} (this is also observed for all other thicknesses explored in this study).

A comparison of Figure~\ref{fig:layered} with Figure~\ref{fig:manufacturable} shows that the effective dose ratios are similar for all three thicknesses, indicating that we are able to reproduce the shielding efficiency of the ideal layered structure by optimizing the manufacturable composite. Similar to the crossover in the optimal number of layers in Figure~\ref{fig:layered}, the optimal manufacturable configuration for $\SI{7}{\mm}$ consists of a trilayer structure at $\omega_{local}$=20\%. As the total thickness of the composite increases to $\SI{100}{\mm}$, the effective dose ratio is minimized when there are five to eight mixture layers ($\omega_{local}$=20\% and areal density of $\sim \SI{10.3}{\gram\per\cm^2}$). This crossover in the manufacturable results indicates that the physical process driving the change in the number of layers is similar to that occurring in the ideal layered geometry. Figure~S2 provides a comprehensive characterization of this crossover feature for the manufacturable configuration. 

Since h${}^{10}$BN is difficult and expensive to manufacture, we run another set of simulations of HDPE interlayered with h${}^{\text{Na}}$BN/HDPE. The results of these simulations, shown in Figure~S3, are compared with the performance of the ${}^{10}$B enriched manufacturable configurations. From this, we find that the effective dose ratio decreases up to a factor of $\sim$ 1.6 when using h${}^{\text{Na}}$BN instead of h${}^{10}$BN within the mixture layers. This decrease in performance varies with energy range and areal density (see Figure~S3). However, it is important to note that, at specific neutron energies (cold, epithermal, intermediate) and large areal densities ($> \SI{5}{\gram\per\cm^2}$), there is not a significant improvement in the shielding efficiency by using $^{10}$B enriched hBN. 

%% ------------------------------------------------------%%

\section{Implication of Geant4 Results}\label{sec:discussion}

\begin{figure*}
\includegraphics[width=1.5\columnwidth]{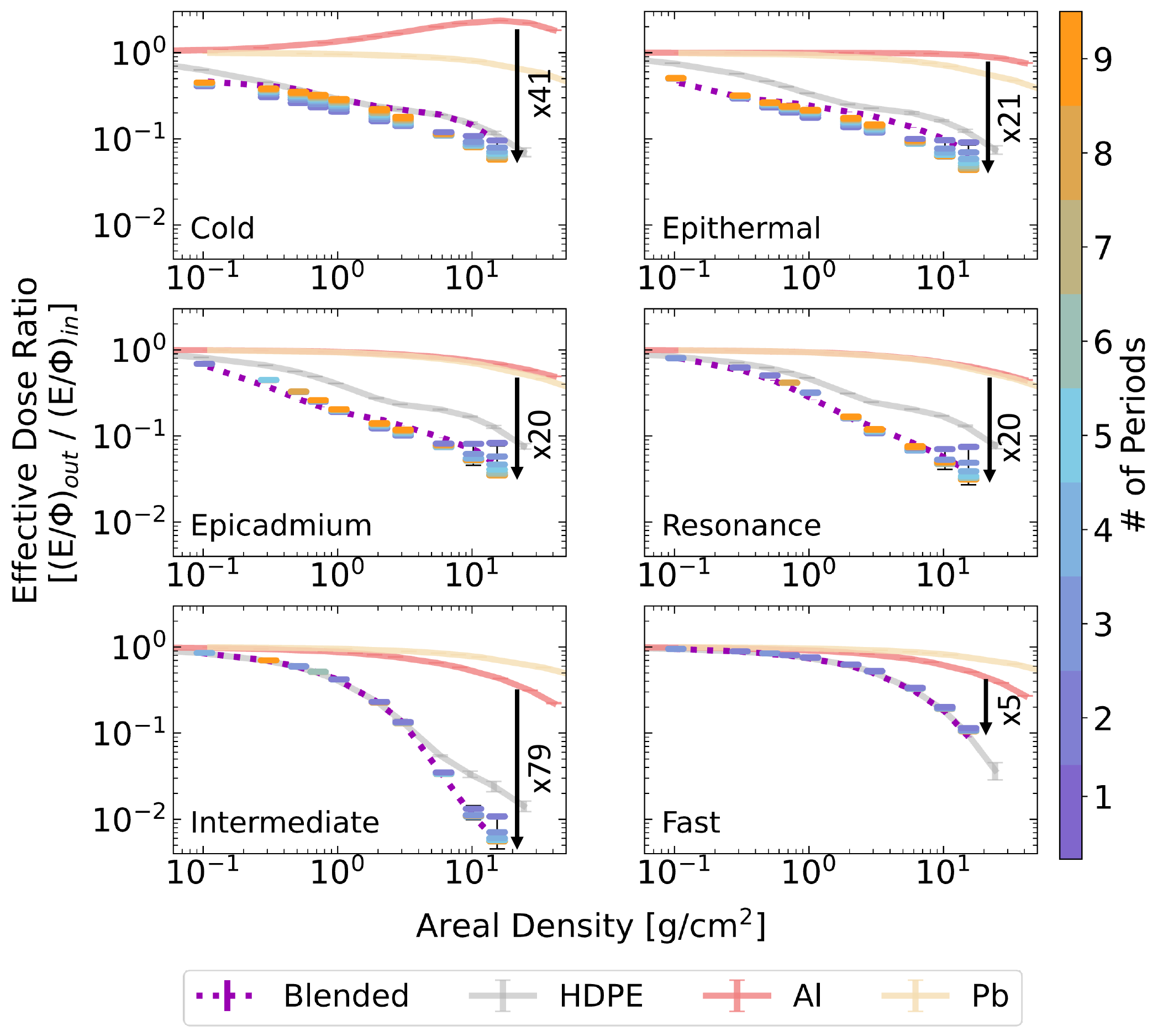}
\caption{\label{fig:summary} Effective dose ratio as a function of areal density for the blended (homogenous blend of HDPE and h${}^{10}$BN) and manufacturable configurations (interlayering of HDPE with h${}^{10}$BN/HDPE). The colorbar shows the different number of periods within the manufacturable configuration. Each panel shows the dose ratio for each energy range explored in this study (shown in Table~\ref{tab:table1}). The reference data for HDPE, Al, and Pb are provided in light gray, coral, and yellow, respectively. Error bars represent the $2\sigma$ error determined by the statistical variation of five random-number seeds, shown in the respective colors for the blended and reference geometries and in black for the manufacturable configuration.}
\end{figure*}

The optimal distribution of $^{10}$B within HDPE can be determined by comparing the performance of the different geometries described in Section \ref{sec:results}. From the MC simulations, we found that separating the thermalization and capture process (ideal layered configuration) provides better shielding than a composite blend, which led us to explore the manufacturable configuration. Figure~\ref{fig:summary} shows the effective dose ratio versus areal density for each energy range in Table~\ref{tab:table1}. The optimal blended configurations, determined by the lowest effective dose ratio for $\omega_{global} \leq 20$\%, are shown as purple dotted lines. The manufacturable configurations are shown for an example of $\omega_{global}$=50\% and $\omega_{local}$=20\%; we note that the results are similar for other values of $\omega_{global}$. For the manufacturable configuration, the colors ranging from blue to orange correspond to the number of periods within each composite, as shown by the colorbar. Reference geometries are shown for comparison: HDPE in light gray, Al in coral, and Pb in yellow. The error bars (generally smaller than the symbol sizes) show the $2\sigma$ variation from running the Geant4 simulations with five different random-number seeds. 

For cold neutrons (E $\leq \SI{0.025}{\eV}$), the composition of the optimal blended and manufacturable configurations changes with areal density. Below $\sim \SI{0.3}{\gram\per\cm^2}$, the optimal blended configurations include small loading of $^{10}$B, demonstrating that it is beneficial to capture the cold neutrons  (also see Figure~\ref{fig:blended}). As the areal density increases, the optimal blended configurations fall along the HDPE curve, indicating that the optimal structure is entirely HDPE. Cold neutrons have a very high probability of undergoing a capture event, and most capture events produce prompt gammas, which also contribute to the effective dose (shown in Figure~\ref{fig:effective_dose}). This tradeoff of cold neutrons for gammas seems to be inefficient for the blended configuration for areal densities above $\sim \SI{0.4}{\gram\per\cm^2}$. However, the manufacturable configuration has a better shielding performance than the blended configurations, reducing the effective dose ratio by an additional factor of 1.12--1.67 across all areal densities. Similar to the blended configurations, the internal composition changes with areal density for the manufacturable configuration, from trilayer to multilayer composite (also shown in Section \ref{sec:manufacturable_results}). This cross-over transition occurs at an areal density of $\sim \SI{5}{\gram\per\cm^2}$ (see Figure~S2 in the supplemental material for a comprehensive characterization of the transition for all energy ranges). For both the blended and layered composites, the optimal configurations reduces the effective dose ratio up to a factor of 41x over Al and 2x over HDPE, which is a significant improvement. 

As the neutron energy increases to the epithermal range ($\SI{0.025}{\eV} <$ E $\leq \SI{0.6}{\eV}$), the optimal blended configuration no longer lies along the reference curve for HDPE, indicating that the inclusion of ${}^{10}$B is important for reducing the effective dose ratio. Similar to cold neutrons, the change in the internal configuration from trilayer to multilayer occurs around an areal density of $\sim \SI{5}{\gram\per\cm^2}$, indicating that the driving physics is likely to be similar for both energy regimes. For areal densities above $\sim \SI{0.1}{\gram\per\cm^2}$, the optimal manufacturable configurations reduce the effective dose more than the blended configurations, resulting in an improvement of up to 1.41x over the blended configurations. At $\sim \SI{0.1}{\gram\per\cm^2}$, the effective dose ratio is lower for the blended composite than a layered composite, indicating that there is no benefit to separating the thermalization and capture mediums. For the composites explored in this study, the optimal configurations reduces the effective dose ratio up to a factor of 21x over Al and 3x over HDPE.

For epicadimum neutrons ($\SI{0.6}{\eV}$ $<$ E $\leq$ $\SI{10}{\eV}$), the optimal blended configurations include \SI{20}{\wt} of h$^{10}$BN for areal densities below $\sim \SI{0.7}{\gram\per\cm^2}$ and between \SI{5}{\wt} and \SI{20}{\wt} for areal densities above $\sim \SI{0.7}{\gram\per\cm^2}$. In contrast to cold and epithermal neutrons, the manufacturable configuration is not the ideal structure for all areal densities --- below $\sim \SI{0.7}{\gram\per\cm^2}$ the blended configuration reduces the effective dose up to 1.20x over the optimal manufacturable configurations, and above this areal density the optimal layered composite reduces the ratio up to 1.30x over the blended. Above $\sim \SI{0.7}{\gram\per\cm^2}$, the optimal manufacturable configuration changes from a trilayer to multilayer configuration around an $\sim \SI{5}{\gram\per\cm^2}$, which is the same cross-over point as the cold and epithermal neutrons. This interesting result shows the importance of optimizing the design space for h$^{10}$BN/HDPE composites as the arrangement of the moderation and capture medium is not obvious and can result in an improvement of 20x and 3x over Al and HDPE, respectively. 

The behaviour of resonance neutrons ($\SI{10}{\eV} < E \leq \SI{300}{\eV}$) is similar to epicadimum neutrons: (1) the blended configurations improve the effective dose ratio up to a factor of 1.19 over the optimal manufacturable configurations at areal densities below $\sim \SI{2}{\gram\per\cm^2}$ and (2) above $\sim \SI{2}{\gram\per\cm^2}$, the structure of the layered composites changes from trilayer to multilayer as the areal density increases. The optimal configurations reduce the effective dose up to a factor of $\sim$ 20x over Al and $\sim$ 4x over HDPE  for resonance neutrons. All three of these energy ranges (epithermal, epicadmium, resonance) show that there is a clear improvement of including h${}^{10}$BN into the HDPE composite. For all the areal densities, the layered composite seems to be the best way to distribute ${}^{10}$B but there are a few instances, at small areal densities, where it is beneficial to have a homogeneous blend of h${}^{10}$BN and HDPE.

For the intermediate range ($\SI{300}{\eV} <$ E $\leq \SI{1}{\MeV}$), the blended and manufacturable composites have the same performance as pure HDPE until the areal density increases to $\sim \SI{3}{\gram\per\cm^2}$. Above $\sim \SI{3}{\gram\per\cm^2}$, the optimal blended or manufacturable configurations have the similar shielding efficiencies, improving the effective dose by 79x over Al and 4x over HDPE. The improvement over Al is larger than the other energy ranges because neutrons are significantly more damaging to human health within this energy range, as shown in Figure~\ref{fig:effective_dose}, resulting in a larger reduction in the effective dose ratio upon thermalizing and capturing these hazardous neutrons. Thus, above $\sim\SI{3}{\gram\per\cm^2}$, there is a benefit to including ${}^{10}$B within the composite as the intermediate neutrons are able to undergo elastic interactions with HDPE and increase their probability of ${}^{10}$B capture events. However, within the range of thicknesses studied, there is no benefit to using a blended or layered composite for this energy range. For fast neutrons ($\SI{1}{\MeV} <$ E $\leq \SI{20}{\MeV}$), there is no benefit to including ${}^{10}$B across all areal densities explored in this study due to the large thickness of HDPE ($> \SI{250}{\mm}$) that is required to thermalize fast neutrons. 

\begin{figure}
\includegraphics[width=1\columnwidth]{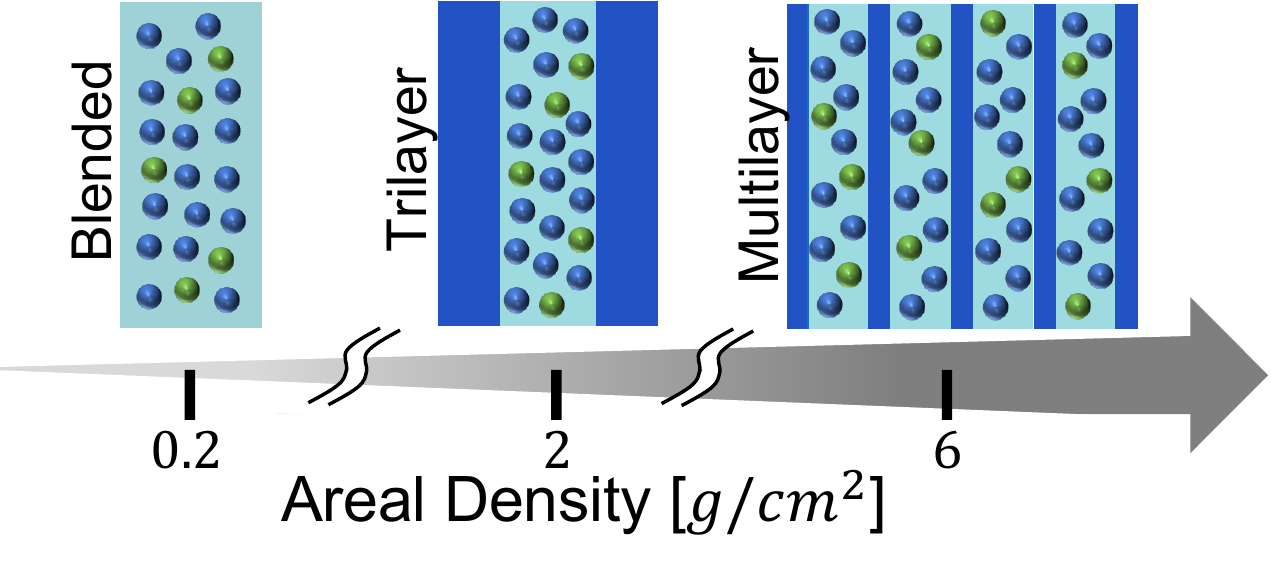}
\caption{\label{fig:summarycartoon} Summary of the change in internal structure for the epicadmium energy range as areal density increases. The blue and green circles provide a representation of HDPE and hBN concentrations, respectively, and the solid dark blue is pure HDPE. The optimal structure changes from blended (below $\sim \SI{0.7}{\gram\per\cm^2}$) to trilayer (between $\sim \SI{0.7}{\gram\per\cm^2}$ and $\sim \SI{5}{\gram\per\cm^2}$) to multilayer (above $\sim \SI{5}{\gram\per\cm^2}$).}
\end{figure}

Across all energy ranges (except for fast neutrons), we discover a dependence where the optimal manufacturable configurations change from a trilayer to a multilayer structure with areal density. This interesting feature could be due to the trade-off of neutrons for gammas and the attenuation of the gammas through the composite. However, there are specific regions where the blended configurations have a better shielding performance than the manufacturable configurations (epicadmium neutrons below $\sim \SI{0.7}{\gram\per\cm^2}$ and resonance below $\sim \SI{2}{\gram\per\cm^2}$). A summary of the change in internal structure for the epicadimum neutrons is shown in Figure \ref{fig:summarycartoon} as a function of areal density. This change in internal structure shows the importance of simulating over the design space of h${}^{10}$BN/HDPE composites as it is experimentally time-consuming to fabricate and test different configurations. Future work involves understanding the underlying physics driving such a change by developing an analytical model. 

We note that there are additional methods to further improve the shielding effectiveness for the materials explored in this study. The effective dose could be further reduced with the addition of high-density material, such as bismuth or barium titanate, at the backend of the composite shielding materials to reduce the gammas that penetrate the shielding material and subsequently contribute to the effective dose. However, the addition of a high-Z material layer to attenuate the photons is outside the scope of this work and could affect the inherent flexibility of the h${}^{10}$BN/HDPE composites. Considering real shielding solutions, the longevity of the hBN/HDPE composites could be affected by accumulated energy from secondary gamma production.

%% ------------------------------------------------------%%
\section{Conclusion}\label{sec:conclusion}

% Original Question: To protect human health, what is the optimal distribution of thermalization and capture elements within a shielding material?

%What have we learned?

%1) For the goal of protecting human health, the addition of boron to HDPE is beneficial. (This may have made sense to folks, but due to the creation of gammas and the different weight factors, the answer wasn't obvious.)

%2) For the same areal density (and also true for the same physical thickness), shielding can be improved beyond a simple blend of materials by choosing an internal structure which isolates the absorption into discrete layers.

%3) As the areal density increases, the optimal layering of absorbing layers with strictly moderating layers changes relatively suddenly from a simple trilayer structure to a multilayer short-period alternation of moderating and absorbing layers. 

%4) Comparisons to aluminum are fine, but aren’t so directly relevant to the point of the study. Comparisons to HDPE are relevant, since it’s the easy option for AM.

In this study, we use Monte Carlo simulations to design the boron distribution within HDPE composites for shielding neutrons to address a fundamental question: what is the optimal distribution of moderation and capture elements within a composite? We explore three different configurations: blended, ideal layered, and manufacturable. We use a log-uniform neutron source from $\SI{15}{meV}$ to $\SI{20}{MeV}$, separated into standard neutron energy ranges, to provide the flexibility to optimize the composite structures for specific neutron applications (e.g. reactor with specific proportions of epithermal and intermediate neutrons). We find that, depending on the incident neutron energies and areal density, the optimal design of hBN/HDPE composites reduce the effective dose ratio by a factor of 5--79x in comparison to Al and 1.5--4x in comparison to HDPE. This is a significant improvement in shielding effectiveness that could dramatically reduce the radiation exposure occupational workers receive. 

From optimizing the internal structure of hBN/HDPE, we discover that the optimal structure changes with areal density and incident neutron energy. In almost all cases, the layered composite reduces the effective dose ratio more than a homogeneous blend. There are some cases where a homogeneous blend is sufficient --- composites with an areal density of $< \SI{0.7}{\gram\per\cm^2}$ and $< \SI{2}{\gram\per\cm^2}$ for epicadmium and resonance neutrons, respectively. For all other cases, the layered composite is the most efficient way to reduce the effective dose ratio. The layered structure changes from a trilayer (two HDPE layers around one blended h${}^{10}$BN/HDPE layer) to a mulitlayer structure as the areal density increases, which is another interesting feature. By exploring a comprehensive design space for hBN/HDPE composites, we learn that the answer to our original question of the best way to distribute moderation and capture elements is nuanced but it is generally beneficial to incorporate h${}^{10}$BN within the HDPE shield. The constraints placed on the manufacturable configuration ensure that the explored design space is compatible with additive manufacturing techniques, a preferred tool for space applications. Moreover, the HDPE and h${}^{10}$BN/HDPE blends used in the manufacturable configuration are inherently flexible, allowing for easy incorporation into the linings of personal protective equipment. Future work involves fabricating and irradiating the optimal manufacturable configurations to confirm shielding effectiveness in real-world composites in addition to incorporating realistic particle distributions into Geant4 simulations. 

%% ------------------------------------------------------%%

\section*{Supplementary Material}

See the supplementary material for Figure~S1, illustrating the interdependence of the design parameters in the manufacturable configuration, Figure~S2, illustrating the transition from trilayer to multilayer for all neutron energy regimes (cold, epithermal, epicadmium, etc.), and Figure~S3, illustrating the improvement factor of using h${}^{10}$BN instead of h${}^{\text{Na}}$BN within the mixture layer of the manufacturable configurations.

%% ------------------------------------------------------%%

\begin{acknowledgments}

The authors acknowledge Dr. Shaheen Dewji for numerous conversations about effective dose and Emma Livernois for detailed editing of the manuscipt. The authors also acknowledge the Georgia Tech Partnership for an Advanced Computing Environment (PACE), which was used to run all the Geant4 simulations. This work was supported by NASA-REVEALS SSERVI (CAN No. NNA17BF68A) and NASA-MSFC (CAN No. 80NSSC21M0271).

\end{acknowledgments}

%% ------------------------------------------------------%%

\section*{Author Declarations}

\textbf{Conflict of Interest} \\

The authors have no conflicts to disclose. \\

\textbf{Author Contributions} \\

\textbf{Alisha Vira}: Conceptualization (lead); software (lead); formal analysis (lead); writing – original draft (lead); writing – review and editing (equal). \textbf{Elizabeth Mone}: Visualization (equal); validation (equal). \textbf{Emily Ryan}: Methodology (equal); writing – review and editing (equal). \textbf{Patrick Connolly}: Validation (lead). \textbf{Karl Smith}: Software (equal); conceptualization (supporting); writing – review and editing (equal). \textbf{Caleb Roecker}: Conceptualization (supporting); software (supporting); writing – review and editing (equal). \textbf{Katherine Mesick}: Conceptualization (supporting); writing – review and editing (equal). \textbf{Thomas Orlando}: Project Administration (lead). \textbf{Zhigang Jiang}: Resources (lead); supervision (equal). \textbf{Phillip First}: Funding Acquisition (lead); supervision (equal); methodology (lead); writing – review and editing (equal).

%% ------------------------------------------------------%%

\section*{Data Availability Statement}

The data that support the findings of this study are openly available in Zenodo at
\href{http://doi.org/10.5281/zenodo.8247757}{http://doi.org/10.5281/zenodo.8247757}.

%% ------------------------------------------------------%%

\appendix

\section{Details of Geant4} \label{app:geant4details}

\begin{figure*}[!t]
\includegraphics[width=1.7\columnwidth]{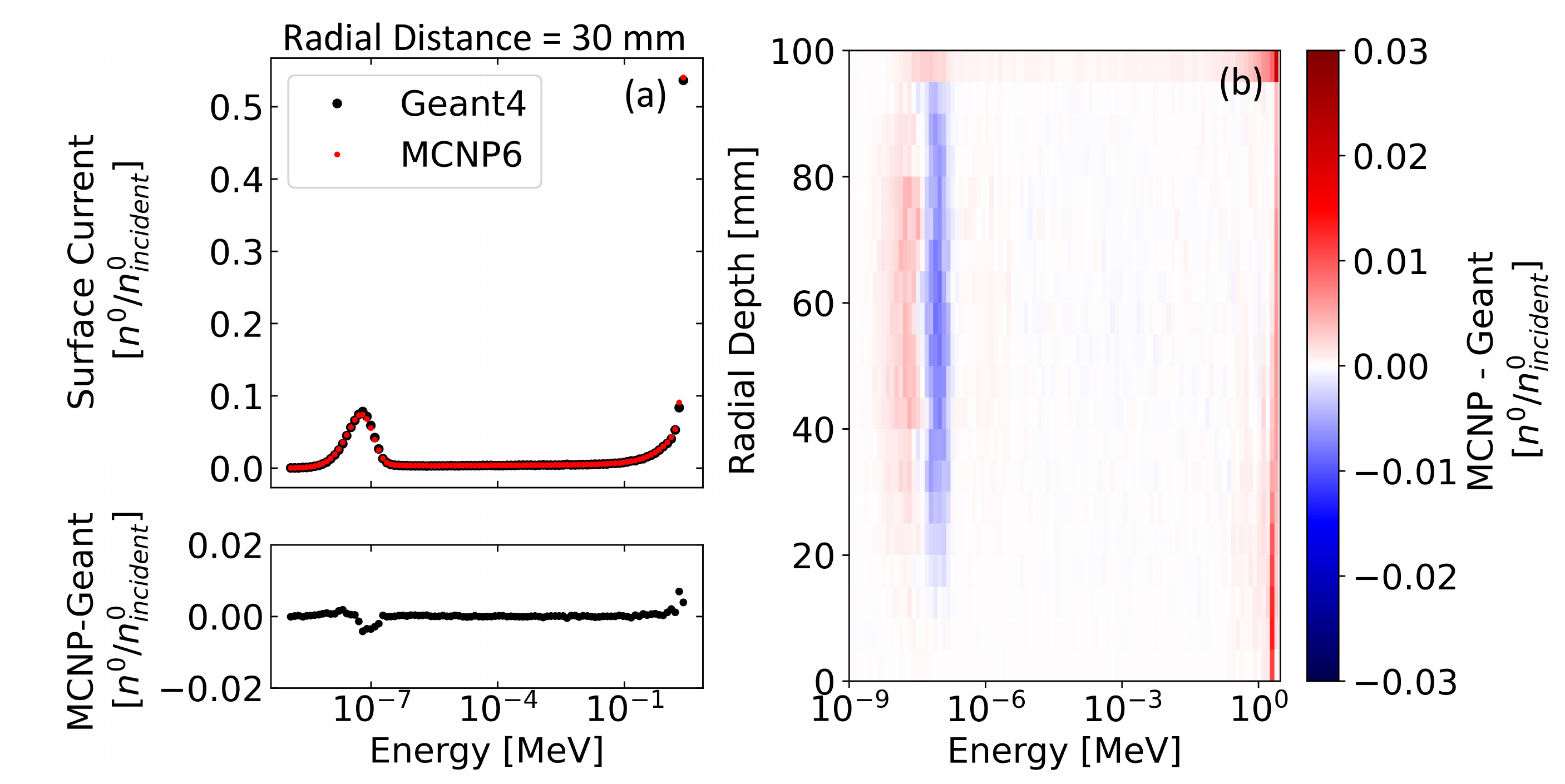}
\caption{\label{fig:MCNPcomparison} (a) Top panel: MCNP (red) and Geant4 (black) neutron surface current comparison at a shell radius of 30 mm from an isotropic source at $\SI{2.5}{\MeV}$ located at the center of an HDPE sphere of radius $\SI{100}{\mm}$. Bottom panel: The difference between MCNP and Geant4 in the top panel. (b) Stacked plot of the difference between MCNP and Geant4 as a function of energy. The difference is plotted on a diverging color bar, where white indicates that the MCNP and Geant4 surface current values are equal. The surface current is normalized to the total number of incident neutrons. }
\end{figure*}

We choose to use Geant4 as it provides additional flexibility that is difficult to achieve in other transport codes. Within Geant4, there are several ways to record the relevant information \cite{agostinelli2003geant4,allison2006geant4,allison2016recent}. The simulation toolkit allows the user to define a region as a “sensitive detector” (SD) and record information related to a “hit”, defined as a physical interaction on a single-particle basis. For each hit, the kinetic energy, charge, and position vector can be recorded, providing the user with detailed information of the particle’s trajectory through a material. This method allows the user to record all the relevant information, then perform post-processing on the simulation results. Alternatively, the user can define specific variables to tally within the code (e.g., deposited energy, surface flux, number of particles), which is useful when there is a specific variable of interest. For the simulations presented in this study, we record the following information associated with each hit: (1) particle type (e.g. neutron, gamma, alpha, etc.), (2) process type (e.g. inelastic collision, elastic collision, ionization, transport, etc.), (3) particle kinetic energy, and (4) particle position in Cartesian coordinates. From this information, the exact interaction and energy deposited into the material can be determined and tracked through the particle cascade. However, to get valid results, it is critical to use the appropriate physics models. For these simulations, we use the following physics constructors:  a modified \textit{G4HadronElasticPhysics} to include \textit{NeutronHPPhysics}, \textit{G4HadronPhysicsShielding}, \textit{G4EmStandardPhysics\_option4}, \textit{G4EmExtraPhysics}, \textit{G4IonPhysics}, \textit{G4DecayPhysics}, \textit{G4RadioactiveDecayPhysics}, and \textit{G4StoppingPhysics}.

Specifically, we modify the original G4HadronElasticPhysics to include the \textit{NeutronHPPhysics} constructor, which computes the phonon density of states for both the coherent and incoherent part and has been verified and found to have a reasonable agreement with other MC simulation codes along with experimental data \cite{thulliez2022improvement,hartling2018effects,lamrabet2021geant4,li2020validation,li2017application}. The \textit{G4HadronElasticPhysics} models the elastic physics of particles, as opposed to the \textit{G4HadronPhysicsShielding}, which models inelastic physics. Both of these constructors are required to obtain accurate descriptions of the interactions occurring for all particles, including thermal neutrons. The other physics constructors are used to account for the following physics: electromagnetic effects (\textit{G4EmStandardPhysics\_option4}, \textit{G4EmExtraPhyics}), ion interactions (\textit{G4IonPhysics}), decay channels according to the branching ratios (\textit{G4DecayPhysics}), radioactive decay of isotopes (\textit{G4RadioactiveDecayPhysics}), and nuclear capture at rest for negatively charged particles (\textit{G4StoppingPhysics}). To ensure that we have loaded the appropriate physics constructors, we compare the Geant4 results to another MC simulation code, Monte Carlo N-Particle code (MCNP), a well-established neutron transport code. 

\section{Validation of Geant4 using MCNP} \label{app:MCNPverfication}

MCNP is a general-purpose radiation tracking code developed by Los Alamos National Laboratory (LANL) for simulating neutrons, photons, electrons, and coupled particle transport \cite{XTeam2003}. It has a wide range of applications, including nuclear reactors, radiation protection, medical physics, etc. In the past 40 years, it has been extensively evaluated and benchmarked \cite{brown2008verification}, providing the standard simulation tool for nuclear engineers. This makes MCNP, specifically MCNP6, an ideal software to validate our Geant4 simulation results. 

To compare the two MC codes, a simulation is run for a simple spherical geometry to remove any ambiguity due to boundary conditions. A solid sphere, placed in a vacuum environment, with a radius of 100 mm is filled with either HDPE or h${}^{\text{Na}}$BN. A monoenergetic neutron source is placed at the center of the sphere and emits neutrons isotropically. For HDPE, we use a density of $\SI{0.968}{\density}$ and a neutron energy of $\SI{2.5}{\MeV}$ so that the thermalization process is observed within the sphere. For h${}^{\text{Na}}$BN, a density of $\SI{2.1}{\density}$ and a neutron energy of $\SI{100}{\keV}$ so that the capture events are observed. For the MCNP simulation, the ENDF/B-VIII.0 cross-section database and thermal scattering card are used. The surface current is tallied at each spherical shell, which is placed at $\SI{5}{\mm}$ increments starting from the center of the sphere and propagating to the surface. 

\subsection{Calculation of Surface Current Tally}

We use the surface current tally, denoted as $F1$ in MCNP, to compare the Geant4 and MCNP simulation results. This tally counts each neutron that crosses a specified surface and weights that neutron by the angle at which it exits the surface. The surface current tally is mathematically represented as
\begin{equation} \label{eq:A1}
F1=\int_A dA \int_E dE \int_{4\pi} d\Omega\ |\hat{n} \cdot \mathbf{J}(r_s,E,\Omega)|.
\end{equation}
where $A$ is the specific surface area, $\hat{n}$ indicates its normal direction, $r_s$ is its distance to the neutron source, $E$ is the neutron energy, and $\mathbf{J}$ is the current vector. By computing the absolute value of $\hat{n} \cdot \mathbf{J}$, MCNP does not distinguish between forward and backward scattering \cite{XTeam2003}. The same calculation is performed in Geant4 during the post-processing of the simulation results. The entire sphere is set as the SD and the direction is weighted by |$\hat{n} \cdot \mathbf{J}|$, replicating the process shown in Equation~\ref{eq:A1}. For simplicity, the $F1$ tally is calculated for neutrons only in both Geant4 and MCNP. 

\subsection{Comparison Results}

Figure~\ref{fig:MCNPcomparison}a (top panel) shows the calculated neutron surface current, $F1$, at a shell radius of $\SI{30}{\mm}$, from MCNP6 (red) and Geant4 (black) for an isotropic source at $\SI{2.5}{\MeV}$ located at the center of an HDPE sphere with a radius of $\SI{100}{\mm}$. Here, MCNP is run for 108 neutrons, and Geant4 is run for 250,000 neutrons. We use a smaller number of neutrons in MCNP as the computational time is longer when recording each individual interaction within the sphere. For direct  comparison, the surface current is normalized by the number of incident neutrons, $n^0_{incident}$. 

The difference between MCNP and Geant4 surface currents, at a shell radius of $\SI{30}{\mm}$, is shown in Figure~\ref{fig:MCNPcomparison}a (bottom panel), where we observe a small discrepancy between the two MC simulation codes. In Figure~\ref{fig:MCNPcomparison}b, the difference between MCNP and Geant4 is shown for each radial shell taken at $\SI{5}{\mm}$ radius increments. The data are plotted with a diverging color bar, where white indicates that the two MC codes give the same results, red indicates that MCNP has a larger surface current than Geant4, and blue indicates that Geant4 has a larger surface current than MCNP. From Figure~\ref{fig:MCNPcomparison}b, we observe that MCNP and Geant4 results are in reasonable agreement with each other for all radial depths within the sphere (within 1\% error between the two codes). After performing the same calculation for an hBN sphere and $\SI{100}{\keV}$ isotropic source, we also find that the MCNP and Geant4 results are in reasonable agreement (within 1\% error), indicating that the physics in Geant4 is appropriate. 

\section{Data Reduction}\label{app:datareduction}

For each configuration, a separate executable is created for the user to specify the $t$, $\omega_{local}$, $\omega_{global}$, and number of layers (not applicable for the blended configuration). The Geant4 code is designed to track the following variables on the input and exit surface of the shielding composite: (1) particle number, type of particle (e.g. neutron, gamma, electron, etc.), type of physics interaction that occurred (i.e. elastic scattering, Compton scattering, etc.), energy of the particle, and the position of a particle ($x$,$y$,$z$). Geant4 outputs these relevant variables to ROOT files, a convenient file format for CERN-developed code \cite{brun1997root}. The post-processing of the Geant4 results is handled in Python. Due to the large parameter size, described in Section \ref{sec:structure}, we utilize PACE, a high performance computing cluster at Georgia Tech, to run $\sim 20,000$ simulations in parallel. 

\begin{figure}
\includegraphics[width=0.7\columnwidth]{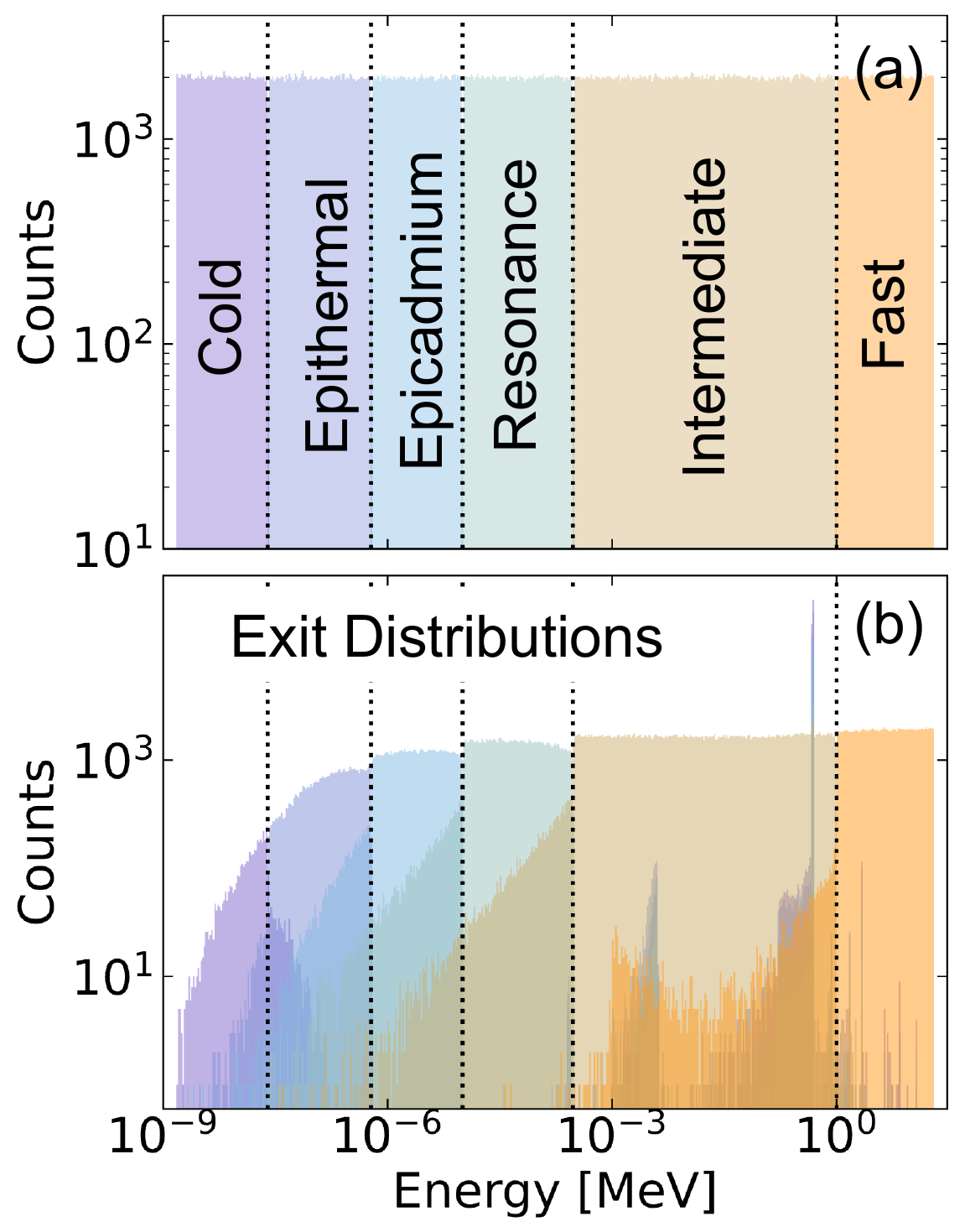}
\caption{\label{fig:dataprocessing} (a) Incident log-uniform distribution separated by neutron energy bins, provided in Table~\ref{tab:table1}. The output distribution for each energy bin for all particles is shown in (b). Both of these subplots are for a blended composite with a total thickness of \SI{1}{\mm} and $\omega_{global}$=10\%. }
\end{figure} 

The ROOT files are read into Python using pyROOT where the ROOT files are converted into \href{https://arrow.apache.org/docs/python/feather.html}{feather files}, a lightweight binary format. All the necessary information for post-processing is saved into the feather files, including the energy of each particle at the entry and exit surface along with the particle type and interaction type. To reduce the overall size of each feather file, the data from $\sim$ 100 ROOT files are saved into a single feather file. The feather files reduce the overall size of the stored data files from $\sim \SI{4}{TB}$ to $\sim \SI{300}{GB}$. This is a significant reduction without losing any information from the ROOT files. 

The feather files are used to create pickle files that contain the final post-processed data. The neutrons are binned by their energy,  shown in Table~\ref{tab:table1}. The binned data is saved to a pickle file, a convenient way to store data in Python \cite{van1995python}. Figure~\ref{fig:dataprocessing}a shows the log-uniform incident neutron source binned by the neutron energy ranges. The distribution at the output surface of the composite is determined using the particle number for each energy range, as shown in Figure~\ref{fig:dataprocessing}b. This final step in the data reduction uses all the data from the feather files to create arrays with the relevant figure of merit. This last step reduces the size from $\sim \SI{300}{GB}$ to $\sim \SI{2}{MB}$, which is a manageable size to work with on a local machine. 

%% ------------------------------------------------------%%

\bibliography{references}% Produces the bibliography via BibTeX.

\end{document}